\documentclass[twocolumn,prl,aps,showpacs]{revtex4}
\usepackage{xspace,amsmath,amsfonts,amsthm,amssymb,amsbsy,graphicx,color}

\begin{document}

\newtheorem{theo}{Theorem}
\newtheorem{lemma}{Lemma}

\title{Rapid State-Reduction of Quantum Systems Using Feedback Control}

\author{Joshua Combes}
\author{Kurt Jacobs}

\affiliation{Centre for Quantum Computer Technology, Centre for Quantum Dynamics, School of Science, Griffith University, Nathan 4111, Australia}

\begin{abstract}
We consider using Hamiltonian feedback control to increase the speed
at which a continuous measurement purifies (reduces) the state of a
quantum system, and thus to increase the speed of the preparation of
pure states. For a measurement of an observable with $N$ equispaced
eigenvalues, we show that there exists a feedback algorithm which
will speed up the rate of state-reduction by at least a factor of
$2(N+1)/3$.
\end{abstract}


\maketitle

To perform information processing using quantum systems the systems
in question must  be prepared in pure states~\cite{QCreview}. Due to
environmental noise such systems  often exist naturally in mixed
states, and either a process of cooling~\cite{Tannor} or measurement
must be used to purify them.  While it is often assumed for the sake
of simplicity that measurements are instantaneous, this is not the
case in reality. All physical processes happen on some finite time
scale, and measurement is no exception. Thus every von Neumann
measurement is in fact a continuous measurement which has been run
for sufficiently long. The speed at which one can measure a system
will therefore place limits on the speed at which measurement can be
used as a tool to prepare initial states.

Given a fixed measurement rate, it is possible to use Hamiltonian
feedback~\cite{B88,YK98,WM93b,DJ99,DHJMT00,MJ04} during the
measurement to increase the speed at which the system is purified.
For a single qubit this problem was analyzed by Jacobs~\cite{J03}.
For this case the optimal feedback algorithm will produce a factor
of 2 increase in the speed at which a given level of purity can be
obtained. This speed-up factor is achieved in the limit in which the
desired purity is high, being the limit in which  the measurement
time is large compared to the measurement rate. Shorter measurement
times result in a smaller speed-up. The speed-up is a quantum
mechanical effect, in  that the equivalent measurement on a
classical bit cannot be enhanced in this  manner; for a two-state
system a speed-up is only possible if the system can exist in a
superposition of the eigenstates of the measured observable during
the process.  Specifically, the optimal algorithm consists of
measuring $\sigma_z$, and using Hamiltonian feedback to continually
rotate the state of the system as needed so that it remains diagonal
in the $\sigma_x$ basis throughout the measurement. This requires a
process of real-time feedback, since the required stream of unitary
operations depends upon the continuous stream of measurement
results.

Here we consider using feedback to enhance a measurement on a system
of arbitrary size. This problem is considerably more complex than
that of a single qubit, because one cannot directly derive the
analytical results which are available in that case~\cite{FJ01}. We
will consider a continuous measurement of an observable with $N$
distinct, equally spaced eigenvalues (i.e. one that distinguishes
the $N$-states), present a feedback algorithm, and prove a lower
bound on its performance. Using this lower bound we show that the
speed-up factor which the algorithm provides in the long time limit
is at least $2(N+1)/3$. Thus for large systems Hamiltonian feedback
can greatly enhance the rate of state-reduction.

The evolution of a system under the continuous measurement of an observable $X$ is given by~\cite{measnote,Bxx,K00,C93,WM93,BS88,D86,B87}
\begin{equation}
  d\rho = -\gamma [X,[X,\rho]] dt + \sqrt{2\gamma}(X\rho + \rho X - 2\langle X\rangle \rho) dW ,
\label{xmeas}
\end{equation}
where $\rho$ is the density matrix of the system and $dW$ is an increment of the Wiener process, describing driving by Gaussian white noise, and $\gamma$ is a positive constant often referred to as the {\em measurement strength}. The continuous stream of measurement results, $y(t)$, is given by $dy(t) = \langle X(t)\rangle dt + dW/\sqrt{8\gamma}$. The effect of such a measurement in the long time limit is to project the system onto one of the eigenstates of $X$, producing a von Neumann measurement in the basis of $X$.

The rate at which the measurement distinguishes between two eigenstates of $X$ is proportional to $\gamma$ multiplied by the difference between the respective eigenvalues. In order to able to purify any initial state the observable must therefore possess non-degenerate eigenvalues. As mentioned above we will also assume that the eigenvalues of $X$ are equally spaced. This is true, for example, if the system has total angular momentum $j$, and the observable is the $z$-component of angular momentum, $J_z$. In this case there are $N = 2j + 1$ states, with eigenvalues $-j,\ldots ,j$. Further, since Eq.(\ref{xmeas}) is invariant under the transformation $X\rightarrow X' = X + \alpha I$ for any real $\alpha$, an overall shift in the eigenvalues of the measured observable has no effect on the measurement. As a result, without loss of generality we may always assume that $X$ is traceless. Under the restriction that the eigenvalues are equally spaced, we may therefore always assume that $X$ is proportional to $J_z$, where $j = (N-1)/2$. While the constant of proportionality is unimportant for the purposes of our analysis, we will define $X$ so that its eigenvalues are separated by $1/N$. Thus $X=J_z/N$.

Using the algorithm in~\cite{J03} as a guide, we will consider a feedback algorithm consisting of a measurement of $X$, in which the feedback is used to rotate the state of the system so that the eigenbasis of the density matrix remains unbiased with respect to the eigenbasis of $X$ during the measurement. Recall that two bases are unbiased with respect to one another when the moduli of the elements of the matrix which transforms between them are all identical~\cite{Wxx}. We will refer to the eigenbasis of $X$ as the measurement basis. To implement such an algorithm one applies a Hamiltonian in each time interval to rotate the state of the system so as to cancel the infinitesimal change in the eigenbasis caused by the measurement in that time interval. In analyzing the performance of this algorithm, we will use the impurity, $L = 1 - \mbox{Tr}[\rho^2]$~\cite{Lnote}, as a our measure of mixedness, rather than the von Neumann entropy, because it has a very simple analytical form. In addition, the impurity is equal to the von Neumann entropy in the limit of high purity.

To begin with we calculate the instantaneous derivative of $L$ when the measurement basis is unbiased with respect to $\rho$. In this case, if we work in the eigenbasis of $\rho$, all the diagonal elements of $X$ are the same, and equal to $\langle X \rangle$, and we have further that $\mbox{Tr}[X^n\rho^m] = \langle X^n \rangle \mbox{Tr}[\rho^m]$. Using this fact in Eq.(\ref{xmeas}) we find that the evolution of $L$ is particularly simple:
\begin{equation}
  dL = -8\gamma Tr[X\rho X\rho] dt .
\label{meas2}
\end{equation}
That is, when the measurement basis is unbiased with respect to $\rho$, the evolution of the impurity over the succeeding infinitesimal time step is deterministic. In the absence of any feedback, the action of the measurement changes the eigenbasis of $\rho$, so that the measurement does not remain unbiased. Under our feedback algorithm, however, Eq.(\ref{meas2}) remains true, and the purity therefore evolves deterministically.

The key to obtaining a lower bound on the performance of our feedback algorithm is the following lemma.

\begin{lemma}
  Consider a traceless observable $X$ of an $N$ dimensional quantum system, whose eigenbasis is unbiased with respect to a density operator $\rho$, and consider the $N!$ unitary operators whose action is to permute the eigenvectors of $\rho$. There exists a set of $N!$ operators $X^{(m)} = U_m X U_m^\dagger$ all unbiased with respect to the state $\rho$ of a system, such that when these operators are measured simultaneously with strength $\gamma$, the instantaneous derivative of $L \equiv 1 - \mbox{Tr}[\rho^2]$ is given by $dL/dt= -8k\gamma N!L$, where $k=\mbox{Tr}[X^2]/(N(N-1))$.
\label{thm1}
\end{lemma}

\begin{proof}
  Consider a simultaneous measurement of $M$ traceless operators $X^{(m)}$ unbiased with respect to $\rho$. From Eq.(\ref{meas2}), the derivative of the impurity is
\begin{eqnarray}
  dL & = & -8\gamma dt \sum_m Tr[X^{(m)}\rho X^{(m)}\rho] \\
           & = & -8\gamma dt \sum_m \sum_{ij} |X_{ij}^{(m)}|^2 \lambda_i \lambda_j ,
\end{eqnarray}
where the $X_{ij}^{(m)}$ are the matrix elements of the $X^{(m)}$ in the eigenbasis of $\rho$, and the $\lambda_i$ are the eigenvalues of $\rho$. Since the $X^{(m)}$ are traceless, $X_{ii}^{(m)} = 0, \; \forall i,m$. Now consider that the $X^{(m)}$ are the $N!$ permutations of a single unbiased traceless operator $X$. In this case, for $i\not= j$ we have $\sum_{m} |X_{ij}^{(m)}|^2 = (N-2)! \sum_{k}\sum_{l\not= k}  |X_{kl}|^2  = (N-2)! \sum_{kl} |X_{kl}|^2 \equiv c$. Naturally when $i=j$ we have $\sum_{m} |X_{ij}^{(m)}|^2 = 0$. As a result
\begin{eqnarray}
  dL & = & -8\gamma dt \sum_{ij} c (1 - \delta_{ij}) \lambda_i \lambda_j \\
           & = & -8\gamma c dt L .
\end{eqnarray}
Permuting the elements of the operator $X$ when written in the eigenbasis of $\rho$ corresponds to a basis change which involves permuting the vectors of the eigenbasis of $\rho$. Such a transformation leaves $X$ unbiased with respect to $\rho$, and thus all the $N!$ operators $X^{(m)}$ are unbiased with respect to $\rho$, and related to $X$ by unitary transformations.

One can also calculate the constant $c$ by using $\sum_m Tr[X^{(m)}\rho X^{(m)}\rho] = c dt L$ and setting $\rho = I/N$. The result is $c = (N-2)!\mbox{Tr}[X^2]$.
\end{proof}

This result shows us two things. The first is that if we measure simultaneously $N!$ unbiased rotated versions of $X$, and combine this with a feedback algorithm which maintains the eigenvectors of $\rho$, the impurity will decay as
\begin{equation}
  L(t) = e^{- 8 \gamma (N-2)! \mbox{\scriptsize Tr}[X^2] t} L(0).
\label{fbrate}
\end{equation}
The second is that, because we know that the derivative of the impurity when summed over all of the $N!$ measurements is $-8\gamma N! k L$, we know that there is at least one term in this sum for which the derivative is greater than or equal to $-8\gamma k L$. That is, given an unbiased operator $X$, and a density matrix $\rho$, we can always find a transformation $T$ (which permutes the eigenbasis of $\rho$), such that measuring $TXT^T$ will provide an instantaneous change in the impurity greater than or equal to $-8\gamma k L$. Of course, we can equivalently apply the transformation to $\rho$, rather than to $X$, which simply involves permuting the eigenvalues of $\rho$.

We can now return to our feedback algorithm. Recall that this involves measuring a single unbiased operator $X$, and applying a unitary operation to the system at each time step to keep the eigenvectors of $\rho$ unchanged. However, using our unitary operation we can also permute the eigenvectors of $\rho$. Therefore, at each time step, we can choose the permutation $T$ so as to maximize the derivative of the impurity. As we have now shown that this maximum derivative is bounded from below by $-8\gamma k L$, we see that our feedback algorithm will produce an impurity $L_{\mbox{\scriptsize fb}}(t)$ such that
\begin{equation}
  L_{\mbox{\scriptsize fb}}(t) \leq e^{- 8 \gamma \mbox{\scriptsize Tr}[X^2] t/(N^2 -N)} L(0).
\end{equation}

We now have a lower bound on the performance of our feedback algorithm. However, what we really wish to know is the enhancement that the algorithm provides over the unaided measurement. In this case the evolution is more complex than that under the feedback algorithm because it is stochastic. However, Eq.(\ref{xmeas}) can be solved using, for example, the technique developed in reference~\cite{JK98}. If the state of the system is initially completely mixed, then the solution can be written, up to a normalization constant, as
\begin{equation}
  \rho(t,v) \propto e^{-4\gamma t (X^2 + 2 v X)} .
\end{equation}
Here $v$ is the measurement record integrated up until time $t$, and divided by $t$. That is,
\begin{equation}
  v(t) = \frac{1}{t}\int_0^t dy = \frac{1}{t}\left[\int_0^t \langle X(s) \rangle ds + \int_0^t
  \frac{dW(s)}{\sqrt{8\gamma}} \right] .
\end{equation}
As will become clear shortly, $v(t)$ is the final result of the measurement after measuring for 
a time $t$: if $v$ is close $n$ at time $t$, and $t \gg N^2/(8\gamma)$, then the system has been very 
nearly projected onto the eigenstate of $X$ with eigenvalue $n$ (recall that $n$ takes the values 
$-j/N,\ldots ,j/N$). The probability density for $v$ is
\begin{equation}
   P(v,t) = \frac{1}{N} \sum_{n=-j/N}^{j/N} \sqrt{\frac{4\gamma t}{\pi}}
            e^{-4\gamma t (v-n)^2} .
\end{equation}
Thus for times long compared to $1/\gamma$, $v$ is sharply peaked about the $N$ values $-j/N,\ldots,j/N$, each peak having width $1/\sqrt{8\gamma t}$.

The density matrix remains diagonal in the eigenbasis of $X$, and its diagonal elements are
\begin{equation}
   \rho_n(t) = e^{-4\gamma t (v-n)^2}/{\cal N} ,
\end{equation}
where the normalization is ${\cal N} = \sum_{n} e^{-4\gamma t (V-n)^2}$. We see that the size of the
$n^{\mbox{\scriptsize th}}$ element is determined by the proximity of $n$ to $v$. When $v$ is close to $n$,
and $t$ is large, $\rho_n$ is much greater than all the other elements.

We wish to calculate the average impurity in the long time limit, so as to compare it with the impurity
achieved under the feedback algorithm. That is, $\langle \bar{P(t)} \rangle = 1 - \langle \mbox{Tr}[\rho^2]
\rangle = 1 - \int_{-\infty}^{\infty}  \mbox{Tr}[\rho^2] P(v) dv$. We will show first that for long times the
average purity, $\langle \mbox{Tr}[\rho^2] \rangle$ for an $N$ state system is bounded from above by that for
a measurement of $\sigma_z/2N$ on a two-state system. Note first that for long times $P(v)$ is sharply peaked
at the $N$ eigenvalues of $X$, so that $\langle\mbox{Tr}[\rho^2]\rangle$ can be written as the average of $N$
integrals, in each of which we only need consider values of $v$ which are close to a single value of $n$.
Since this is true for every $N$, the only difference between the average purity for different values of $N$
comes from the purity $\mbox{Tr}[\rho^2]$. Now consider the two integrals corresponding to the extreme values
of $n$ ($n=j/N$ and $n=-j/N$). We note that for any value of $N$, if we truncate the density matrix so that we
keep only the two largest elements (and renormalize), then the value of $\mbox{Tr}[\rho^2]$ in these integrals
becomes precisely that describing a measurement of $\sigma_z/2N$ on a two state system. The factor of $1/N$
stems from the fact that adjacent  eigenvalues of $X$ are separated by $1/N$. Further, the process of
truncation increases the purity. This follows immediately from the theory of majorization applied to the
diagonal elements of $\rho$, and the fact that the purity is a symmetric convex function of these
elements~\cite{Bhatia}. Thus, the two integrals corresponding to the extreme values of $n$ are bounded from
above by those for a measurement of $\sigma_z/2N$. Further, it is immediately clear that for every $N$, the
value of $\mbox{Tr}[\rho^2]$ under the $(N-2)$ integrals corresponding to the non-extreme values of $n$, upon
truncation, give a smaller purity than those for the extreme values (this is because for these values of $n$,
there are adjacent values on {\em both} sides, and not merely one side). Since the total average purity is
simply the average value of the $N$ integrals, for every $N$ this average is bounded from above by that for a
measurement of $\sigma_z/2N$. Thus for long times the value of $L$ for such a measurement provides a lower
bound on a measurement of $X$ for $N>2$. (In fact, numerical results show that this is true at all times -- as
indeed one would intuitively expect -- and not merely in the limit in which $t\gg N^2/(8\gamma)$.)

The average value of the impurity for a measurement of $\sigma_z /2N$ may be written as~\cite{J03}
\begin{equation}
   L_2(t) = \frac{e^{-\gamma t/N^2}}{\sqrt{8\pi t}} \int_{-\infty}^{+\infty} \frac{e^{-x^2/(2t)}}{\cosh(\sqrt{2\gamma}x/N)} dx .
\end{equation}
Noting that the integral in this expression is independent of $t$ for $t \gg N^2/(8\gamma)$, we have, in this limit, $P_2(t) = e^{-\gamma t/N^2}C/(\sqrt{8\pi t})$, where $C = \int_{-\infty}^\infty 1/\cosh(\sqrt{2\gamma}x/N) dx $.

Now that we have an upper bound on $L$ under our feedback algorithm, and a lower bound on this for the unassisted measurement, we can derive a lower bound on the speed-up factor, $S$, provided by the algorithm. Specifically, $S$ is the ratio between the time taken to achieve a given target value of the purity under the unaided measurement, $t_{\mbox{\scriptsize m}}$, and that required when using the algorithm, $t_{\mbox{\scriptsize fb}}$. The result is
\begin{equation}
 \frac{1}{S} \leq \frac{1}{8N^2k}
                  + \frac{\ln [\tilde{C} \sqrt{t_{\mbox{\scriptsize m}}}]}
                         {8k\gamma t_{\mbox{\scriptsize m}}} ,
\end{equation}
where $\tilde{C} = \sqrt{8\pi}(N-1)/(CN)$. Using $k = \mbox{Tr}[X^2]/(N(N-1))$, evaluating $\mbox{Tr}[X^2] = \mbox{Tr}[J_z^2]/N^2 = \sum_{n=-j}^{j} n^2 = (N+1)(N-1)/(12N)$, and taking the limit of large $t$, we obtain
\begin{equation}
 S \geq \frac{2}{3}(N+1).
\end{equation}

\begin{figure}
\leavevmode\includegraphics[width=0.95\hsize]{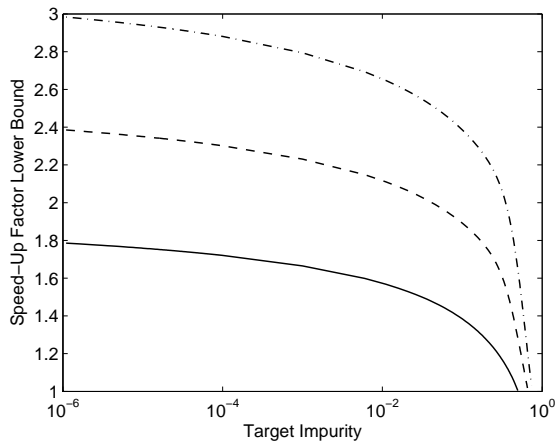}
\caption{We plot here a lower bound on the speed-up factor provided
by our feedback algorithm as a function of the target impurity, when
the initial state is completely mixed. The solid line gives the
result for system dimension $N=2$, the dashed line for $N=3$ and the
dot-dashed line for $N=4$.} \label{fig1}
\end{figure}

This expression is valid in the limit of high target purity. We have also calculated numerically the lower bound on the speedup factor as a function of the target purity to examine how it approaches the above limit, and we plot the results for $N=2,3$ and $4$ in figure~\ref{fig1}. We find that this lower bound increases monotonically towards its limiting value as the target purity is increased.

Before we finish we describe in a little more detail the feedback Hamiltonian involved in implementing the algorithm. To implement the feedback algorithm the controller is required to apply a unitary operation to the system so as to preserve the basis in which the density matrix, $\rho$, is diagonal, and, if necessary, apply a unitary operation to reorder the eigenvalues of $\rho$. To calculate the first unitary the controller diagonalizes $d\rho$. Since $d\rho$ has two terms, one proportional to $dt$ and the other to $dW$, this unitary will in general take the form $U(t,dt) = e^{i(H_1(t)dt + H_2(t)dW)}$ for some operators $H_1$ and $H_2$. The feedback Hamiltonian is thus, in general, $H(t) = H_1(t) + H_2(t)(dW/dt)$, and therefore contains a term proportional to the noise part of the measurement record $y(t)$. This kind of quantum feedback has been studied extensively by Wiseman and Milburn~\cite{WM93}. The second unitary, that required to reorder the eigenvalues of $\rho$ should ideally perform the operation in a single time step. This requires a Hamiltonian proportional to $1/dt$, and since $dt$ should ideally be small on the time scale of the dynamics due to the measurement, this would require a Hamiltonian large compared to $\gamma$.

A number of interesting open questions remain. While we have found a lower bound on the performance of our algorithm, its actual performance may be higher. In addition, we do not know whether this algorithm is optimal. Further, in the above analysis we have focused on the ability of feedback to increase the speed of state-reduction under the assumption that there is no significant limitation on the feedback Hamiltonian. It will be interesting to consider the performance of the algorithm when specific limits are placed on the feedback Hamiltonian, and this will be the subject of future work.

{\em Acknowledgements:} The authors would like to thank Tanmoy Bhattacharya, Julia Kempe and Howard Wiseman for helpful discussions.


\begin{thebibliography}{99}
\bibitem{measnote}
  An introduction to continuous quantum measurement is given in~\cite{Bxx}.
  Other recent treatments of continuous measurements of Hermitian operators are
  given in~\cite{DJ99,K00}, where the first uses the results of optical continuous
  measurements in~\cite{C93,WM93}. A selection of the original work on continuous
  measurement is given in references~\cite{B88,BS88,D86,B87}.
\end{thebibliography}
\end{document}